\documentstyle[twocolumn,12pt,epsf]{article}
\title{\normalsize\bf A General Signal of a
Phase Transition \\
from Single-Particle Momentum Distributions }

\begin{document}

\author{\normalsize H. Q. Wang  \\
     \normalsize Department of Cosmic and Subatomic Physics, \\
     \normalsize  University of Lund, Sweden \\}

\date{\small\flushright A two-particle space correlation function
is derived from the single-particle momentum distribution of \\ 
the emission source. A signal of a first order phase transition is 
obtained from this correlation function if \\
\ \ density fluctuations are large. \ \ \ \ \ \ \ \ \ \ \ \ \ \ 
\ \ \ \ \ \ \ \ \ \ \ \ \ \ \ \ \ \ \ \ \ \ \ \ \ \ \ \ \ \ \ \ \ \ \
\ \ \ \ \ \ \ \ \ \ \ \ \ \ \ \ \ \ \ \ \ \ \ \ \ \ \ \ \ \ \ \ \ \ \
\ \ \ \ \ \ \ \ \ \ \ \ \ \ \ \ \ \ \    \\
\vspace{0.5cm}
{\hspace{-8.2cm} PACS number: 25.75.GZ, 12.38.Mh, 24.60.-k, 64.60.-i}}

\vspace{-1.0cm}  

\maketitle

\small
%\footnotesize

The search for phase transitions is a hot topic in atomic and nuclear
physics. Several different phase transitions are observed/or expected 
to be observed like $^3$He to $^4$He superfluid, Bose-Einstein 
condensation, liquid to gas nuclei and hadronic matter to Quark Gluon 
Plasma (QGP). The QGP phase transition is predicted by lattice QCD and 
several QGP signals have been suggested\cite{signal}, although no 
convincing evidence\cite{QM95} has yet been presented. The nature of the 
phase transition of QGP depends on the dynamical quarks. A first-order 
phase transition is expected without dynamical quarks. Simulations of 
lattice QCD with dynamical quarks have not yet overcome the limitations 
due to finite lattice size. Present results indicate a smooth cross-over 
between phases for two light quark flavours ($N_f$=2) and a first-order 
phase transition for $N_f>$3\cite{signal}.
 
Among these QGP signals there are two based on the study 
of particle average transverse momentum ($\langle p_T\rangle$) and on two 
particle momentum correlations. If the dependence of $\langle p_T\rangle$ 
on 
the (pseudo-) rapidity density ($dn/d\eta$) of produced particles\cite{van} 
is studied, the phase transition of QGP to hadronic matter 
could be seen as a plateau followed by a second rise in $\langle p_T\rangle$. 
However this signal is not sensitive to the shape of the single-particle 
momentum distribution. When studying the emission source space-time extension 
through Bose-Einstein correlations, it was argued that the two particle 
momentum correlations can signal a QGP phase transition\cite{HBTphase} 
through a very large apparent source radius in the 'outward' direction.

Recently much attention has been paid to search for disoriented chiral
condensates (dcc) from single-particle momentum 
distributions and two-particle momentum correlations\cite{gavin}. Bjorken 
et al. pointed out that large fluctuations in the pion spectra
can be expected from the dcc \cite{bjo} which create large variation 
in the source density. 

The fluctuation of the source density near critical point was first 
discussed by Ornstein and Zernike\cite{lsz}.
In this paper we study the two-particle space correlations of the 
emission source through density fluctuations in heavy ion physics. 
The two-particle space correlation function is derived from the 
single-particle momentum distributions. A signal of a first order 
phase transitions sensitive to the shape of the single particle 
momentum distribution is proposed. This signal is not only suitable 
for searching of the QGP
phase transition but also for the search of the liquid to 
gas phase transition in the excited nucleus.

Assume a source at rest with volume $V$ in which there are $N$ particles
distributed in a set of positions $\vec{r}_i$, $i=1,2,...N$. The 
normalized $N$ particle probability distribution  
is $p(\vec{r}_1,\vec{r}_2,...,\vec{r}_N)$. The particle number
density at a position $\vec{r}$ is defined as,  
\begin{equation}
\rho_1(\vec{r}) = \sum_{i=1}^{N}{\delta(\vec{r} - \vec{r}_i)}. 
\end{equation}
Here $\delta$ represents a Dirac delta function. 
Thus $\rho_1(\vec{r})d\vec{r}$ is the number of particles between
$\vec{r}$ - $\vec{r}+d\vec{r}$.
The integration of the source volume will give 
\begin{equation}
\int_{V}\rho_1(\vec{r}) d\vec{r} = N.
\end{equation}
The ensemble average of the particle number density 
$\overline{\rho_1(\vec{r})}$ is assumed to be independent on the position 
$\vec{r}$ and denoted as $\rho$, 
\begin{eqnarray}
\rho&=&\int d\vec{r}_1 ...\int d\vec{r}_N \sum_{i=1}^{N}
\delta(\vec{r}-\vec{r}_i)p(\vec{r}_1,\vec{r}_2,...,\vec{r}_N) \nonumber \\
&=&N\int d\vec{r}_1 \delta(\vec{r} -\vec{r}_1)P(\vec{r}_1) \nonumber \\
                 &=&N \bf  P(\vec{r}).
\end{eqnarray}
The $\rho_1(\vec{r})$ is the local particle density which fluctuates 
around the $\rho$ and $\rho_1(\vec{r})-\rho$ reflects
the density fluctuation of the source.
The volume integral of $\rho$ gives the average number of 
particles $\langle N\rangle$  
\begin{equation}
\int d\vec{r} \rho(\vec{r})=\langle N\rangle.
\end{equation}
The two-particle density correlation function can be expressed by, 
\begin{equation}
\rho_2(\vec{r},\vec{r'})=\sum_{i\neq j=1}
\delta(\vec{r}-\vec{r}_i)\delta(\vec{r'}-\vec{r}_j).
\end{equation}
The ensemble average of $\rho_2(\vec{r},\vec{r'})$ is obtained 
in a similar way as $\rho(\vec{r})$,
\begin{equation}
\overline{\rho_2(\vec{r},\vec{r'})}=N(N-1)P(\vec{r},\vec{r'}).
\end{equation}
The integral of $\overline{\rho_2(\vec{r},\vec{r'})}$ is the mean 
number of two particle pairs $\langle N(N-1)\rangle$ in the volume.

The density-density correlation is defined as 
$\overline{\rho_1(\vec{r})\rho_1(\vec{r'})}$. From the formulas above 
it is found 
\begin{equation}
\overline{\rho_1(\vec{r})\rho_1(\vec{r'})}=\rho \delta(\vec{r}-\vec{r'})+
\overline{\rho_2(\vec{r},\vec{r'})}.
\end{equation}
One can now study the density-density fluctuations from the
density-density correlation,
\begin{equation}
\overline{[\rho_1(\vec{r})-\rho][\rho_1(\vec{r'})-\rho]}=\rho 
\delta(\vec{r}-\vec{r'})+\overline{\rho_2(\vec{r},\vec{r'})}-\rho^{2}.
\label{eq:dr}
\end{equation} 
The Fourier transformation of $\rho_1(\vec{r})-\rho$, is,
\begin{equation}
\rho_1(\vec{r})-\rho=(2\pi)^{-3}\int{\rho_1(\vec{k})\exp(i\vec{k}\cdot 
\vec{r})d\vec{k}}. 
\label{eq:fou}
\end{equation}
Where $\rho_1(\vec{k})$ is the Fourier component of particle number
density fluctuation which is a complex variable.
Because $\rho_1(\vec{r})-\rho$ is real, $\rho_1^{*}(\vec{k})$ = 
$\rho_1(-\vec{k})$ and because of the normalization 
$\rho_1(\vec{r})$, $\rho_1(\vec{k})|_{\vec{k}=0}=0$. Equations \ref{eq:dr} 
and \ref{eq:fou} give
\begin{equation}
\overline{|\rho_1(\vec{k})|^2} =V\rho+V\int e^{-i\vec{k} \cdot \vec{r}}
(\overline{\rho_2(\vec{r})}-\rho^2) d\vec{r},
\end{equation}
by assuming that the two particle space correlation only depends on the 
distance between two particles.
The statistical average on the left side could be found in the
following way. The total free energy $F$ is expressed in terms of 
the free-energy density $F_1(\vec{r})$,
\begin{equation}
F=\int F_1(\vec{r})d\vec{r}.
\end{equation}
For a given temperature, $F_1(\vec{r})$ may deviate from 
its equilibrium value $\overline{F_1}$ due to the density fluctuations. 
Because $F_1(\vec{r})$ 
should be minimum for equilibrium, Let us expand the difference 
$F_1(\vec{r})-\overline{F_1}$ according to Landau's approach\cite{LAN}:
\begin{equation}
F_1(\vec{r})-\overline{F_1} = \frac{1}{2} a[\rho_1(\vec{r})-\rho]^2 
+\frac{1}{2} b(\nabla \rho_1(\vec{r}))^2+...
\end{equation}
here parameters $a$ and $b$ are independent of the density but may 
depend on the temperature. 
For a finite size source the Fourier 
integral should be replaced by the Fourier series. Neglecting 
higher-order terms in eq.(12), 
\begin{equation}
F-\overline{F_1}V=\frac{1}{2V} \sum(a+b\vec{k}^2)|\rho_1(\vec{k})|^2.
\end{equation}
Based on this expression, the probability of having the fluctuation 
$|\rho_1(\vec{k})|$ is given by, 
\begin{equation}
p(|\rho_1(\vec{k})|)=A \exp(-(a+b\vec{k}^2)|\rho_1(\vec{k})|^2/2VT).
\end{equation}
Here
$A=(\int e^{-(a+b\vec{k}^2)|\rho_1(\vec{k})|^2/2VT} d\rho_1(\vec{k}))^{-1}$ 
which implies,
\begin{equation}
\overline{|\rho_1(\vec{k})|^2}=\frac{VT}{a+b{\vec{k}}^2}. 
\end{equation}
Using this and taking the Fourier inverse in eq.(10), we obtain
two particle space correlation function
\begin{equation}
\overline{\rho_2(r)}-\rho^2+\rho\delta(r)=\frac{T}{4\pi b r}
\exp[-(\frac{a}{b})^{1/2}r] \label{eq:phas}
\end{equation}
The parameter $a$ must vanish at the phase transition point 
(critical point $T=T_c$) with zero net-baryon density, because
\begin{equation}
a=(\frac{\partial^2F_1(r)}{\partial\rho_1(r)^2})_T=\frac{1}{\rho_1(r)}
(\frac{\partial P}{\partial \rho_1(r)})_T.
\end{equation}  
Here $P$ is the pressure of the system. For the first-order phase
transition, the two-particle space correlation function becomes long-range,  
\begin{equation}
\overline{\rho_2(r)}-\rho^2+\rho\delta(\vec{r})=
\frac{T_c}{4\pi b r},     \ \ \ \ T=T_c.
\end{equation}
In the vicinity of $T_c$, $a$ is non-zero. Thus 
the two-particle space correlation function decreases exponentially.

Eq.(16) shows that two-particle space correlation function 
contains information of the first order phase transition. Now 
the way of obtaining such information experimentally will be presented 
in accordance with the second quantisation method.

Assume that particles with $\vec{p}=\hbar \vec{k}$ move freely in a static
source of volume $V$ at freeze-out.
The normalized single-particle wave function is 
\begin{equation}
\phi_{\vec{k}}(\vec{r})=\frac{1}{\sqrt{V}}\exp(i\vec{k}\cdot\vec{r}).
\end{equation}
The operators $\hat{a}^+_{\vec{k}}$ and $\hat{a}_{\vec{k}}$ increase and 
decrease in units of the numbers $n_{\vec{k}}$ of particles in the various 
quantum states $\phi_{\vec{k}}$. The operators become,
\begin{equation}
\hat{\Psi}^+(\vec{r})=\sum \phi^*_{\vec{k}}(\vec{r})
\hat{a}^+_{\vec{k}}, \ \ \
\hat{\Psi}(\vec{r})=\sum \phi_{\vec{k}}(\vec{r}) \hat{a}_{\vec{k}}
\end{equation}
which respectively add and remove one particle from point 
$\vec{r}$ in the system. The operator 
$\hat{\Psi^+}(\vec{r}) \hat{\Psi}(\vec{r})$dV is the operator 
of the number of particles in the volume dV. 
Hence $\hat{\Psi}^+ \hat{\Psi}$ 
can be regarded as an operator $\hat{n}$, which represents the particle 
density distribution in space 
\begin{equation}
\hat{n}=\hat{\Psi}^+(\vec{r})\hat{\Psi}(\vec{r})=
 \sum_{\vec{k}} \sum_{\vec{k}^{'}}\hat{a}^+_{\vec{k}}  
\hat{a}_{\vec{k}^{'}}\phi^*_{\vec{k}}\phi_{\vec{k}^{'}}.
\end{equation}
The diagonal terms of the sum ($\vec{k}=\vec{k}^{'}$) give the 
mean density $\overline {n}$ in the quantum state considered,
\begin{eqnarray}
{\overline n}&=&<\vec{k}|\sum_{\vec{k}}{\hat{a}^{+}_{\vec{k}}\hat{a}_{\vec{k}}|
\phi_{\vec{k}}|^2}|\vec{k}>\nonumber \\
&=&\frac{1}{V} \sum {<\vec{k}|\hat{n}_{\vec{k}}|\vec{k}>}=\frac{1}{V} 
\sum n_{\vec{k}}.
\end{eqnarray}
Here operator $\hat{n}_{\vec{k}}=\hat{a}^+_{\vec{k}}\hat{a}_{\vec{k}}$ and 
$n_{\vec{k}}$ is the number of particles in the quantum state. 
The operator which contributes to the density fluctuation is
\begin{equation}
\hat{n}-{\overline n}=\sum_{\vec{k}} \sum_{\vec{k}^{'}\neq \vec{k}}
\hat{a}^+_{\vec{k}}\hat{a}_{\vec{k}^{'}} \phi^*_{\vec{k}}
\phi_{\vec{k}^{'}}.
\end{equation}
The average density-density 
fluctuation is marked as $\overline{(\hat{n}(\vec{r}_1)-
{\overline n})(\hat{n}(\vec{r}_2)-{\overline n})}$. 
This mean value is calculated in two stages. First of all, the quantum 
averaging over all the quantum 
states of $\phi_{\vec{k}}$ and $\phi_{\vec{k}^{'}}$ then the ensember 
(statistics) averaging should be carried out. The average quantum 
state is as
\begin{eqnarray}
&&D2 \nonumber \\
&=&\langle\vec{k}_1\vec{k'}_1\vec{k}_2\vec{k'}_2|(\hat{n}(\vec{r}_1)-
{\overline n})(\hat{n}(\vec{r}_2)-{\overline n})
|\vec{k}_1\vec{k'}_1\vec{k}_2\vec{k'}_2\rangle \nonumber \\
&=&\langle\vec{k}_1\vec{k'}_1\vec{k}_2\vec{k'}_2|\sum_{\vec{k}_1}
\sum_{\vec{k'}_1}^{'}\sum_{\vec{k}_2} \sum_{\vec{k}^{'}_2}^{'}
\hat{a}^+_{\vec{k}_1}\hat{a}_{\vec{k'}_1}\hat{a}^+_{\vec{k}_2}
\hat{a}_{\vec{k'}_2} \nonumber \\
& &\phi^*_{\vec{k}_1}(\vec{r}_1)
\phi_{\vec{k'}_1}(\vec{r}_1)
\phi^*_{\vec{k}_2}(\vec{r}_2)\phi_{\vec{k'}_2}(\vec{r}_2)
|\vec{k}_1\vec{k'}_1\vec{k}_2\vec{k'}_2\rangle.
\end{eqnarray}
$D2$ is not zero although 
$\vec{k}_1=\vec{k'}_2$ and $\vec{k'}_1=\vec{k}_2$. Let $\vec{k}=\vec{k}_1$ 
and $\vec{k'}=\vec{k}_2$, then the average quantum state of eq.(24) 
will become,
\begin{equation}
D2=\frac{1}{V^2}\sum_{\vec{k}}\sum_{\vec{k'}}(1\pm  n_{\vec{k'}})
n_{\vec{k}}\ e^{i(\vec{k}-\vec{k'})\cdot (\vec{r}_1-\vec{r}_2)}.
\end{equation}
Here $n_{\vec{k}}$ is the number of particles in the quantum state. 
The '+' refers to the case of Bose statistics and '-' to that of Fermi
statistics. After the quantum averaging the $D2$ must also be averaged
in the statistical sense, i.e. over the equilibrium distribution of the
particles in the various quantum states. Since particles in 
different quantum states are quite independent, the
numbers $n_{\vec{k}}$ and $n_{\vec{k'}}$ are averaged independently,
\begin{equation}
\overline {(1 \pm n_{\vec{k'}})n_{\vec{k}}}\approx (1 \pm 
{\overline n}_{\vec{k'}}){\overline n}_{\vec{k}}.
\end{equation}
If one now replace the summation in eq.(25) into integration, we obtain 
the following expression for the mean density-density fluctuation,
\begin{eqnarray}
& &{\overline{(\hat{n}(\vec{r}_1)-{\overline n})(\hat{n}(\vec{r}_2)-{\overline n})}} \nonumber \\
&=&\frac{1}{(2\pi)^3}\int{\overline n}_{\vec{k}}\ e^{-i\vec{k}\cdot
(\vec{r}_1-\vec{r}_2)}
\delta(\vec{r}_1-\vec{r}_2)d\vec{k}  \nonumber \\
&\pm &\frac{1}{(2\pi)^6} \int \int {\overline n}_{\vec{k'}}{\overline n}_{\vec{k}}\ e^{i(\vec{k}-\vec{k'})\cdot
(\vec{r}_2-\vec{r}_1)} d\vec{k} d{\vec{k'}}  \nonumber \\
&=&{\overline n}\delta(\vec{r})\pm 
|\frac{1}{(2\pi)^3}\int{\overline n}_{\vec{k}}\ e^{i\vec{k}\cdot\vec{r}}
d\vec{k}|^2,
\end{eqnarray}
here $\vec{r}=\vec{r}_1-\vec{r}_2$ and 
${\overline n}_{\vec{k}}=\frac{1}{e^{(\varepsilon-\mu)/T}\mp 1}$
is the Bose/Fermi distribution.

After comparison of eq.(27) with eq.(8) the two-particle space 
correlation function is found to be,
\begin{equation}
\overline{\rho_2(\vec{r}_1,\vec{r}_2)}-\rho^2=\pm |\frac{1}{(2\pi)^3}\int 
{\overline n}_{\vec{k}}e^{i\vec{k}\cdot\vec{r}}d\vec{k}|^2.
\end{equation}

For a classical (Boltzmann) system, the  $\hat{a}_{\vec{k}}\hat{a}^+_{\vec{k}}
=\hat{a}^+_{\vec{k}}\hat{a}_{\vec{k}}$, so that, 
\begin{eqnarray}
& &{\overline {(\hat{n}(\vec{r}_1)-{\overline n})(\hat{n}(\vec{r}_2)-{\overline n})}} 
\nonumber \\
&=&\frac{1}{(2\pi)^6}|\int {\overline n}_{\vec{k}}
e^{i\vec{k}\cdot\vec{r}}d\vec{k}|^2,
\end{eqnarray}
and the two particle space correlation function is 
\begin{equation}
\overline{\rho_2(\vec{r}_1,\vec{r}_2)}-\rho^2+\rho\delta(\vec{r}_1-\vec{r}_2)=
\frac{1}{(2\pi)^6}|\int{\overline n}_{\vec{k}}e^{i\vec{k}\cdot\vec{r}}
d\vec{k}|^2. 
\label{eq:ct}
\end{equation}
Here ${\overline n_{\vec{k}}}=\exp((\mu-\varepsilon)/T)$ is the 
Boltzmann distribution.

If the spin $S$ of the particles is taken into account, the 
two particle space correlation function will become 
\begin{equation}
\overline{\rho_2(\vec{r}_1,\vec{r}_2)}-\rho^2=\pm\frac{g^2}{(2\pi)^6}  
|\int {\overline n}_{\vec{k},\sigma}e^{i\vec{k}\cdot\vec{r}} d\vec{k}|^2.
\end{equation}
Here $\sigma$ is the spin projection and $g=2S+1$ is the degenerace of the particle. ${\overline n}_{\vec{k},\sigma}$ is the mean number of particles 
at a state of given spin projection $\sigma$.

We must of course be aware of that the difference in momentum between 
two Bosons tend to be small and more pairs are observed than for Fermions.

Equation (28) also shows that the difference between Bosons and Fermions
is similar in configuration space, the presence of one Boson at some point 
increases the probability that another Boson is close to that point.
  
Comparing the experimental distribution of two-particle space correlation
calculated from eq. 28 or \ref{eq:ct} with that from eq. \ref{eq:phas}, 
we could probe whether the first order phase transition has taken place. 
If no clasical density fluctuations and no quantum correlations, the 
particles will be emitted independently in space. Otherwise the 
correlation function becomes long range if particles are emitted 
at the critical point. 

Figure 1 shows the dependence of the two particle space correlation on
the distance between two particles ($r$) with and without a first order 
phase transition. Curve 1 represents the two-particle space 
correlation including the phase transistion ($a=0$) and curve 2 
without the phase transistion 
($a/b\approx 4$). The 
temperature $T$=160 MeV is chosen. 
The calculations show that if parameter $a$ is of the same magnitude 
as $b$, the difference of the two-particle emission pattern from that 
with the phase transition is clearly visible. Two-particle emission
with small separation in distance dominates if there is no phase 
transition. The results are also calculated with a pure 
Boltzmann distribution with $T=$160 MeV and Bose
distributions with two different temperatures. These results show
no signal of a phase transition as expected. 
A possible single-particle momentum distribution (strong
enhancement at low momentum and an enhancement at high momentum) 
which results in a long range correlation of two particles in space is 
also indicated in the figure.
\begin{figure}[bht]
\vspace{-0.8cm}
\epsfxsize=8.cm
\epsfysize=5.0cm
\hspace{0.5cm}
\epsffile{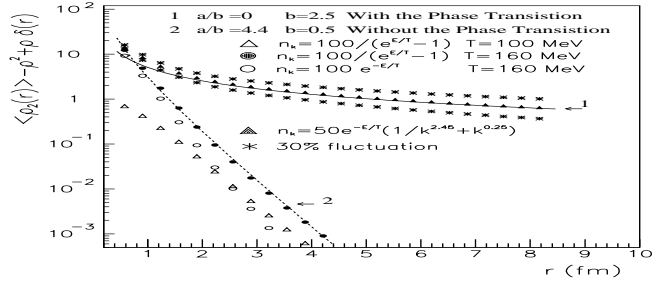}
\vspace{-1.0cm}
\caption{The dependence of two-particle space correlation function on the 
separation distance between two particles.}
\end{figure}
Experimentally, one should search for the phase 
transition on an event-by-event basis, looking for significant 
fluctuations in a single-particle momentum distribution. Assuming that
there is 30\% fluctuation of the magnitude in the fitted single-particle
momentum distribution at momentum less than 30 MeV/c due to detector 
resolution but keeping the function of the momentum distribution the same as 
that indicated by the solid triangle in the figure, the two-particle 
space correlation function will vary and the varied region is also 
indicated in the figure.  

Figure 2 shows the single-particle momentum distributions which are used to 
calculate the two-particle space correlation. It demonstrates that what kind of
shape of single-particle momentum distribution possibly indicate a phase transition. 
\begin{figure}[bht]
\vspace{-0.8cm}
\epsfxsize=8.cm
\epsfysize=5.0cm
\hspace{0.5cm}
\epsffile{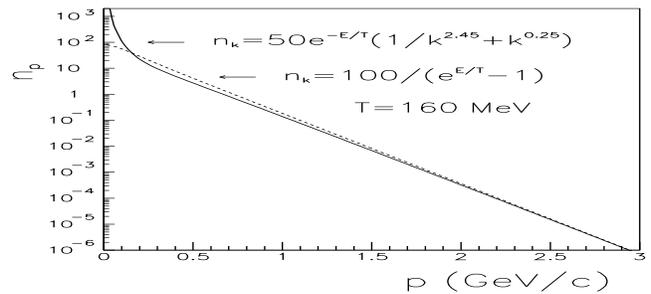}
\vspace{-1.0cm}
\caption{Comparison of two kind of single-particle momentum distributions used 
in the calculation of two-particle space correlation.}
\end{figure}

In eq.(26) the assumption of independent particles 
in different quantum states simplified the calculation of 
the two-particle space correlation. In fact this approximation can be 
removed and the two-particle space correlation can be deduced from 
single-particle and two-particle momentum distributions. \\

\noindent
{\large\bf Acknowledgements}\\
The author is greatly indebted to Dr. B. Jakobsson. He also would 
like to thank the discussions with Professors U. Heinz, 
X. Cai and Drs. S. Garpman and E. Stenlund.

%\noindent
%{\large\bf Figure Captions}\\
  
%\noindent
%{\bf Figure 1:} The dependence of two-particle space correlation %function
%on the separation distance between two particles.

%\noindent
%{\bf Figure 2:} Comparison of two kind of single-particle momentum 
%distributions used in the calculation of two-particle space correlation.

\end{document}